\documentclass[
    aip,
    apl,
    reprint,
]{revtex4-1}

\usepackage{graphicx}
\graphicspath{ {figures/} }

\DeclareUnicodeCharacter{2212}{-}
\DeclareUnicodeCharacter{03B8}{$\theta$}

\usepackage{braket}
\usepackage{amsmath}
\usepackage{amssymb}
\usepackage{xcolor}
\usepackage{hyperref}

\begin{document}

\title{Quantum Kernel Machine Learning for Autonomous Materials Science} 

\author{Felix Adams}
\email[]{fadams@umd.edu}
\affiliation{Department of Materials Science and Engineering, University of Maryland, College Park, Maryland, 20742, USA}

\author{Daiwei Zhu}
\affiliation{IonQ, 4505 Campus Dr, College Park, Maryland, 20740, USA}

\author{David W. Steuerman}
\affiliation{IonQ, 4505 Campus Dr, College Park, Maryland, 20740, USA}

\author{A. Gilad Kusne}
\affiliation{Department of Materials Science and Engineering, University of Maryland, College Park, Maryland, 20742, USA}
\affiliation{Materials Measurement Science Division, National Institute of Standards and Technology, Gaithersburg, Maryland, 20899, USA}

\author{Ichiro Takeuchi}
\affiliation{Department of Materials Science and Engineering, University of Maryland, College Park, Maryland, 20742, USA}
\affiliation{Quantum Materials Center, University of Maryland, College Park, Maryland, 20742, USA}

\date{\today}

\begin{abstract}

Autonomous materials science, where active learning is used to navigate large compositional phase space, has emerged as a powerful vehicle to rapidly explore new materials. A crucial aspect of autonomous materials science is exploring new materials using as little data as possible. Gaussian process-based active learning allows effective charting of multi-dimensional parameter space with a limited number of training data, and thus is a common algorithmic choice for autonomous materials science. An integral part of the autonomous workflow is the application of kernel functions for quantifying similarities among measured data points. A recent theoretical breakthrough has shown that quantum kernel models can achieve similar performance with less training data than classical kernel models. This signals the possible advantage of applying quantum kernel machine learning to autonomous materials discovery. In this work, we compare quantum and classical kernels for their utility in sequential phase space navigation for autonomous materials science. Specifically, we compute a quantum kernel and several classical kernels for x-ray diffraction patterns taken from an Fe-Ga-Pd ternary composition spread library. We conduct our study on both IonQ’s Aria trapped ion quantum computer hardware and the corresponding classical noisy simulator. We experimentally verify that a quantum kernel model can outperform some classical kernel models. The results highlight the potential of quantum kernel machine learning methods for accelerating materials discovery and suggest complex x-ray diffraction data is a candidate for robust quantum kernel model advantage.

\end{abstract}

\maketitle 

\section{Introduction}

Discovery and optimization of new materials play a crucial role in advancing technology in modern society. However, the space of possible compositions and synthesis parameters is immense. Researchers are increasingly relying on high-throughput experimental and computational approaches to identify new functional compounds for their use in a variety of technological areas. Autonomous materials science has emerged as an effective vehicle for navigating the space of possible materials.

Autonomous experimental methods can self-navigate multi-dimensional parameter space through continuous updating of a Bayesian surrogate model without having to survey every point in the parameter space. There have been a number of demonstrations of closed-loop autonomous workflows for successful optimization of compositional parameters (with respect to specific materials characteristics) with only a limited number of experimental iterations in self-guided sequences \cite{kusne_--fly_2020, seifrid_autonomous_2022, stach_autonomous_2021, abolhasani_rise_2023, szymanski_autonomous_2023}.

Combinatorial libraries, where large compositional variations are laid out on individual wafers/chips, are particularly useful physical platforms for materials optimization. Autonomous optimization within libraries can be used to quickly chart materials within fixed composition variations such as ternary phase diagrams \cite{green_applications_2013}.

Figure \ref{fig:active_learning} shows a schematic of an example typical active learning workflow as applied to the Fe-Ga-Pd ternary composition space. The goal of the autonomous workflow is to determine the distribution of the different crystalline phases throughout the composition space (``phase map''), obtained using XRD measurements, using as few measurements as possible. The iterative workflow has 4 steps: measurement, clustering, extrapolation, and decision. First, an XRD measurement is performed at a particular composition. Second, all the XRD patterns measured so far are clustered into groups (representing different crystal phases). Third, the clusters are then extrapolated to the unmeasured compositions. Fourth, the predictions and their uncertainties are used to determine which composition should be measured next. To minimize the total uncertainty in the predicted phase map, the composition with the largest uncertainty is chosen at each iteration. Once the total uncertainty falls below a threshold, we have a reasonable prediction of the phase map of the entire composition space after only measuring a fraction of the composition space.

Autonomous experiments require a machine learning surrogate model for the property of interest. Popular methods such as deep learning require lots of training data, which aren't available for novel materials. The machine learning subfield of Few-Shot Learning (FSL) studies methods for such data-limited regimes \cite{parnami_learning_2022}. Experimental materials science data are expensive and therefore limited in general, making some FSL techniques such as transfer learning (where knowledge from one data-rich task is transferred to another data-limited task) challenging \cite{parnami_learning_2022}.

Gaussian processes are a kernel machine learning method commonly used for the extrapolation step of the autonomous materials science workflow. Gaussian processes are an extension of Gaussian distributions: a Gaussian distribution describes the probability distribution of a random value, while a Gaussian process describes the probability distribution of a function \cite{rasmussen_gaussian_2008}. A multivariate Gaussian distribution has a covariance matrix describing the relations between variables, and a Gaussian process has a covariance function describing the relations between function values. This covariance function is the kernel function of the Gaussian process. For more information on Gaussian processes and Gaussian process classification, see reference \hspace{0 pt} \cite{rasmussen_gaussian_2008}. We used Gaussian process classification as implemented in GPflow \cite{matthews_gpflow_2017}.

Quantum kernel machine learning is an emerging area of research that may enable certain functions to be learned with less training data than classical kernel methods, making it a promising tool for autonomous materials discovery. Recent work by Huang et al. demonstrated that classical kernel models can predict the output of quantum kernel models if they are given sufficient training data \cite{huang_power_2021}. However, we are especially interested in the data-limited regime where classical kernel models don't have enough data to outperform quantum kernel models, such as in the first few iterations of the autonomous materials science workflow. As highlighted by the well-known "No Free Lunch" Theorems \cite{wolpert_no_1997}, the performance of a machine learning model fundamentally depends on the specific problem it is applied to. Therefore, the value of quantum kernel methods hinges on identifying problem domains where they offer a genuine advantage. 

Diffraction data might be such a domain where quantum kernel methods have an advantage. Scattering and diffraction (using X-ray, electrons, and neutrons) are ubiquitous characterization tools in materials science as they provide reciprocal representations of periodic atomic positions as well as their electronic and magnetic structures \cite{kittel_introduction_2004}. Diffraction data might provide a data platform on which quantum kernel models can learn faster than classical kernel models because the techniques used to represent diffraction (matrix inversion and the Fourier transform) are thought to be faster on quantum computers. The transformation from real space to reciprocal space can be written as a matrix inversion \cite{kittel_introduction_2004}, which is an application where quantum computers have an exponential advantage using the Harrow Hassidim Lloyd (HHL) algorithm \cite{harrow_quantum_2009}. While we don’t use either the HHL algorithm or the quantum Fourier transform in this work, the centrality of those problems to analysis of diffraction data suggests that diffraction datasets are a fruitful area to explore for a quantum kernel model advantage.

In this work, we evaluate how well a quantum kernel model can learn an x-ray diffraction (XRD) dataset using IonQ's Aria trapped ion quantum computer (see NIST Disclaimer). We introduce the XRD dataset and a corresponding typical autonomous materials science workflow, describe the classical and quantum kernels used, explain how we compared the kernels, show the results of the comparison, and discuss their implications. Lourenço et al. recently performed similar work with other, non diffraction materials science datasets: perovskite properties and doped nanoparticle relaxation \cite{lourenco_exploring_2024}.

This work presents a novel application of quantum kernel machine learning and demonstrates how recent developments in our theoretical understanding of quantum advantage in machine learning can be fruitfully applied to address scientific and engineering challenges.

\section{Methods}

\subsection{XRD Dataset}

Our experimental materials science dataset is from a ternary Fe-Ga-Pd thin film composition spread which was fabricated by co-sputtering Fe, Fe\textsubscript{2}Ga\textsubscript{3}, and Pd on a 3 inch (7.62 cm) Si wafer such that a continuous variation in composition is created on the wafer \cite{long_rapid_2007}. Such composition-spread combinatorial libraries provide a useful platform for exploring new materials within a large ternary compositional phase diagram. The dataset consists of stoichiometric composition information, i.e. $a$, $b$, and $c$ in Fe\textsubscript{a}Ga\textsubscript{b}Pd\textsubscript{c} (where $a+b+c=1$), and XRD diffraction patterns taken at 237 uniformly distributed positions/compositions on the library.

XRD of the films was performed using the $\omega$-scan mode of a D8 DISCOVER for combinatorial screening (Bruker-AXS) using an x-ray beam spot size of 1 mm diameter. It is equipped with a GADD two-dimensional detector which captures data for a fixed range of 2$\theta$ and $\omega$ at once. The raw detector images are then integrated to obtain the 2$\theta$ angles and peak intensities using the D8 GADDS program and a script to automate the process. For more details on the synthesis and XRD characterization processes, see references \hspace{0 pt} \cite{long_rapid_2007} and \hspace{0 pt} \cite{takeuchi_identification_2003}.

We've used this dataset as a benchmark for other machine learning studies (such as in reference \hspace{0 pt} \cite{kusne_high-throughput_2015}), so the quantum kernel model's performance on this dataset is likely representative of its performance on other materials science datasets. 

We sampled this dataset to simulate the first few measurements made in the autonomous materials science workflow. We used 4 XRD patterns from each of the 5 distinct structural phase regions on the wafer for a total of 20 data points. The ground truth phase labels come from human expert analysis of the XRD data \cite{long_rapid_2007}. 

\begin{figure*}[p]
    \centering
    \includegraphics[width=17cm]{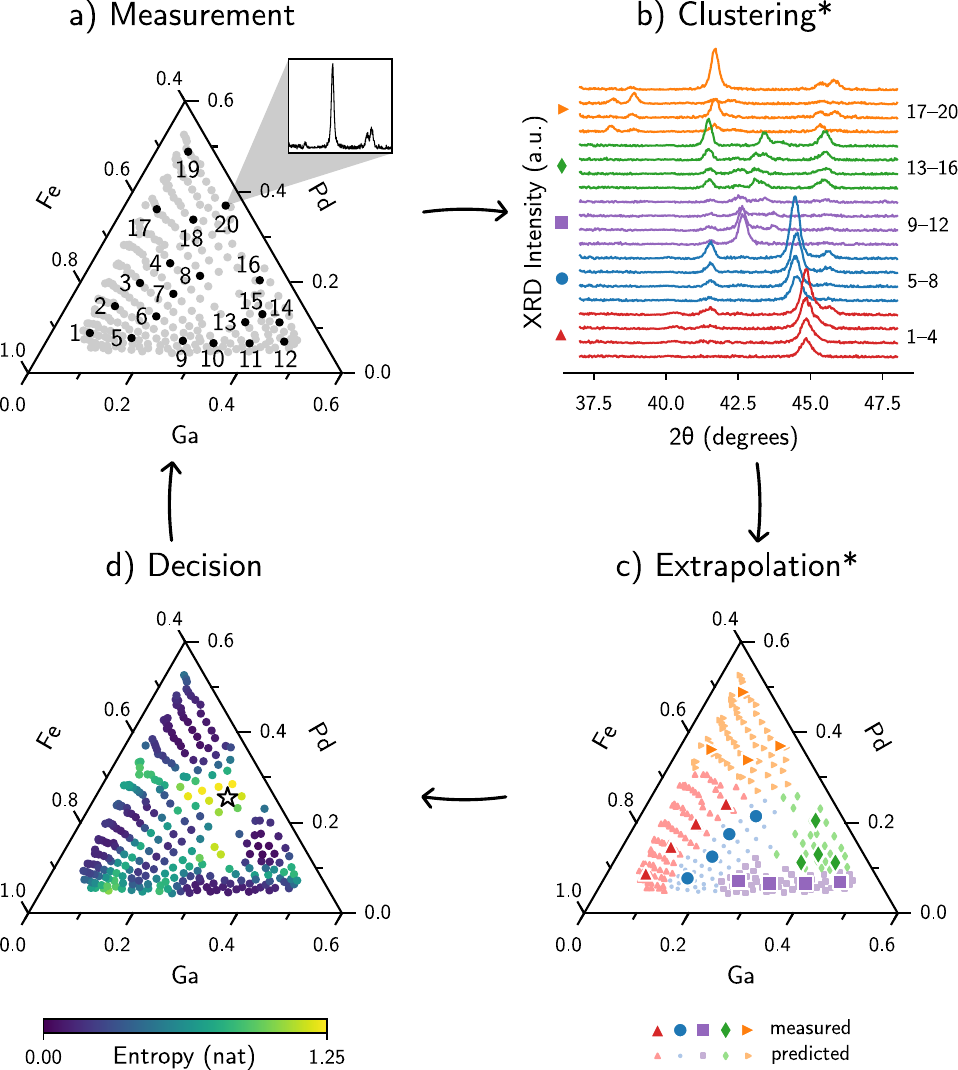}
    \caption{
        A schematic of a typical iterative autonomous phase mapping workflow. Asterisks (*) indicate which steps use kernel methods.
        \textbf{(a)} A new x-ray diffraction measurement is performed at a specific composition (shown in units of atomic fraction). Sample indices 1 -- 20 were chosen manually here for illustration and do not reflect a typical measurement order. 
        \textbf{(b)} All of the measured compositions are clustered into groups and assigned a corresponding label (color and symbol to the left) using an \emph{x-ray diffraction} kernel function. 
        \textbf{(c)} The labels of the unmeasured compositions are then extrapolated using a \emph{composition} kernel function. 
        \textbf{(d)} To minimize the total uncertainty in the extrapolated labels, the composition with the maximum uncertainty (entropy) in its predicted label is chosen as the next composition to measure, indicated by the star. 
        \textbf{Note} that the clustering and extrapolation methods are both kernel methods but use different kernel functions which compare different types of data (x-ray diffraction patterns and compositions).}
    \label{fig:active_learning}
\end{figure*} 

\newpage

\subsection{Classical and Quantum Kernels for Autonomous Materials Science}

Kernel machine learning methods find patterns in data using a kernel function, $k(x_1, x_2)$, which takes two data points $x_1$ and $x_2$ as inputs and returns a value describing how similar the two data points are \cite{mengoni_kernel_2019}. We compared the quantum kernel to two classical kernels, a radial basis function ($k_{\text{RBF}}$) and the cosine similarity ($k_{\text{CS}}$):

\begin{equation}
    k_{\text{RBF}}(x_1, x_2)=\exp{\left(-\frac{\Vert\hat{x}_1 - \hat{x}_2\Vert^2}{2}\right)}
\end{equation}
\begin{equation}
    k_{\text{CS}}(x_1, x_2)=\hat{x}_1 \cdot \hat{x}_2
\end{equation}

...where $\cdot$ is the dot product, $\Vert \hspace{5 pt} \Vert$ is the Euclidean $L^2$ norm, and $\hat{x}_i$ is the $L^2$ normalized data vector $x_i$. In this work, each $x_i$ is a vector containing the XRD intensity values of an XRD pattern.

The two classical kernels presented for comparison were chosen to illustrate the nature of the potential quantum kernel advantage. The squared exponential $L^2$ norm radial basis function kernel is a common kernel for Gaussian processes, capable of describing arbitrary functions which are sufficiently smooth \cite{rasmussen_gaussian_2008}. The cosine similarity is a simpler similarity measure which works well for clustering XRD patterns using other machine learning methods compared to more sophisticated kernels such as dynamic time warping or earth mover's distance\cite{iwasaki_comparison_2017}, but is atypical as a Gaussian process kernel because of its simplicity. The polynomial kernel is common in the machine learning literature for high-dimensional classification\cite{rasmussen_gaussian_2008}, but is less common for analysis of XRD patterns.

These kernels typically have hyperparameters which are fitted to the data during model training. In this work, to simplify analysis and comparison across different training datasets and dataset sizes, we use fixed hyperparameters which worked well for this dataset. 

To calculate a kernel function, $k_Q$, using a quantum computer, we use a feature map quantum circuit, $U$, which receives the values of a data point $x \in \mathbb{R}^d$ ($d$ is the number of data dimensions) as parameters and calculate...
\begin{equation}
    k_Q(x_1, x_2)=\lvert \bra{0}U^{\dag}(x_2)U(x_1)\ket{0} \rvert^2
    \label{kQ}
\end{equation}
...i.e., starting with a set of qubits in the state $\ket{0}$, one applies the feature map circuit with the data point $x_1$, applies the conjugate transpose (i.e, the inverse, because the evolution of a closed quantum system is a unitary transformation\cite{nielsen_quantum_2010}) of the feature map circuit with the data point $x_2$, then finds the probability that the final state of all the qubits remains $\ket{0}$ \cite{havlicek_supervised_2019}. If $x_1$ and $x_2$ are similar, the two feature map circuits mostly cancel out, the probability that the circuit output remains $\ket{0}$ is large, and the kernel value is high. Otherwise, the two feature maps don't cancel out, the circuit outputs an arbitrary state with low probability of being $\ket{0}$, and the kernel value is low. To estimate $k_Q(x_1, x_2)$, the circuit calculation is repeated 1,024 times to determine the approximate probability of measuring $\ket{0}$. This calculation is repeated for every pair of data points to construct a kernel matrix, $K_{i,j}=k(x_i, x_j)$, which is then used by the classical kernel machine learning method. 

We used a feature map quantum kernel circuit based on the work of Peters et al. \cite{peters_machine_2021} to compare XRD patterns. The quantum feature map circuit is shown in Figure \ref{fig:feature_map}. The circuit is composed of quantum operations, or ``gates'', which affect 1 or 2 qubits at a time. The gray, 1-qubit gates are Hadamard gates and the gray 2-qubit gates are $\sqrt{i\text{SWAP}}$ gates. The white, 1-qubit gates are rotation gates, which accept a parameter that determines the angle of rotation. Each group of three rotations (around the $z$, $y$, and $z$ axes again respectively), can describe an arbitrary rotation of the qubit Bloch sphere. The XRD intensities are used as the angle parameters of the rotation gates, after being re-normalized and scaled between $[0, \pi]$ radians. The numbering of the rotation gates indicates the order in which the XRD intensities were used. We used 25 qubits and 6 rotation gates per qubit, totaling 150 XRD intensities (from 150 evenly spaced 2$\theta$ values) used in the feature map circuit (the dimension of the input data vector $x_i$). In this work, we used the IonQ Aria trapped ion quantum computer for the quantum circuit execution. Quantum circuits corresponding to kernel function evaluations comparing any element to itself were not performed, instead assuming the ideal value of 1. The classical simulations of the quantum kernel output were performed using IonQ API with the standard Aria noise model. 

Eq. \ref{kQ} defines not a single quantum kernel, but a family of kernels parametrized by the choice of feature map circuit $U$. In principle, $U$ can be tailored to known structure in the data, such as permutation-equivariant circuits for graph learning tasks \cite{skolik_equivariant_2023} or quantum convolutional architectures for translationally symmetric data \cite{chinzei_splitting_2024}. In our XRD classification problem, however, the absolute positions and relative intensities of diffraction peaks are physically meaningful, so neither global permutations nor rigid translations of the 150 intensity values preserve the labels. As a first step, we therefore focus on a single, hardware-efficient feature map based on Peters et al. \cite{peters_machine_2021} that serves as a generic but expressive ansatz compatible with the IonQ Aria device. 

Quantum algorithms for accelerating kernel machine learning calculations have been developed \cite{rebentrost_quantum_2014}, but require the use of Quantum Random Access Memory \cite{mengoni_kernel_2019}, which is beyond current quantum computers. In this work we used quantum computers to calculate kernel functions which are then used by classical machine learning methods \cite{schuld_quantum_2019, havlicek_supervised_2019}, as this approach is feasible with current technology.

\begin{figure}[!t]
    \centering
    \includegraphics[width=8.5cm]{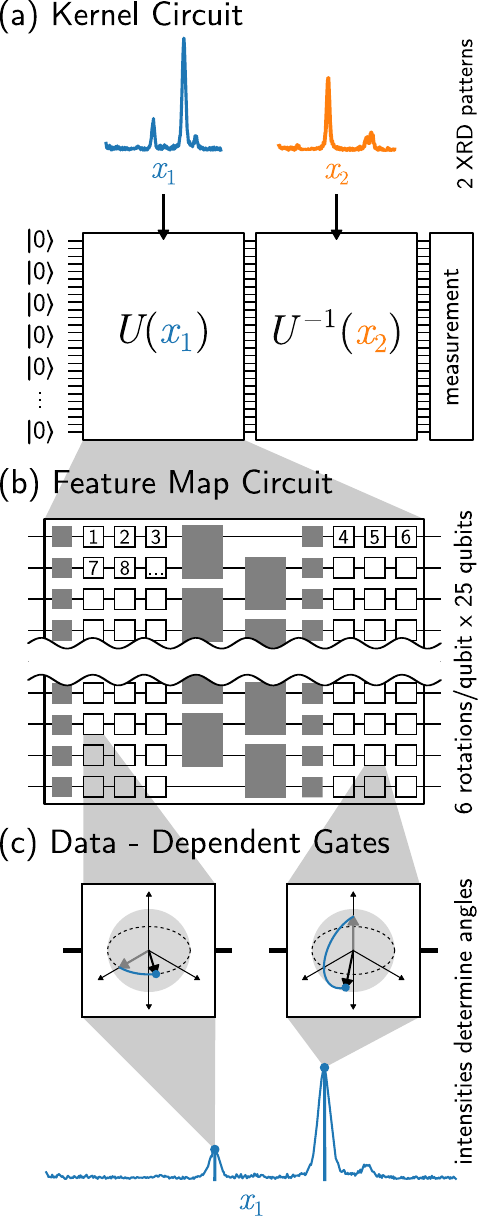}
    
    \vspace{1cm}
\end{figure}

\begin{figure}[t]
    \caption{
        The quantum circuit used to compare two XRD patterns, $x_1$ and $x_2$.
        \textbf{(a)} The kernel circuit is composed of a forward and a reverse copy of the feature map circuit, $U$, which takes the intensities of an XRD pattern as parameters. The initial state of each qubit is $\ket{0}$ and the kernel value is the probability that the final state of every qubit is $\ket{0}$. 
        \textbf{(b)} The feature map circuit, $U$, is composed of 1- and 2-qubit operations, or ``gates''. Each horizontal line represents a qubit, and the circuit computation proceeds from left to right. The gray, 1-qubit gate is a Hadamard gate, the gray 2-qubit gate is a $\sqrt{i\text{SWAP}}$ gate, and the triplets of white gates are rotation gates about the $z$, $y$, and $z$ axes respectively. The numbers indicate the order in which XRD intensities parameterize the rotation gates.
        \textbf{(c)} A schematic illustrating how the XRD intensities determine the angle of rotation in the rotation gates, a $z$ axis rotation on the left and a $y$ axis rotation on the right.
    } 
    \label{fig:feature_map}
\end{figure}

\newpage

\subsection{Evaluating Quantum Kernel Performance}

To determine a quantum kernel's theoretical potential for quantum advantage, one can use the quantities defined by Huang et al.\cite{huang_power_2021}: the model complexities of the quantum and classical kernels and the geometric difference between the kernels.
The model complexity $s$ for kernel matrix $K$ is defined as...
\begin{equation}
    \centering
    s_K(N) = \sum_{i,j=1}^{N}(K^{-1})_{i,j}(2 y_i \cdot y_j - 1)
\end{equation}
...where $K^{-1}$ is the inverse of the kernel matrix, $N$ is the number of data points, and $y_i$ and $y_j$ are the one-hot encodings of the phase labels of each composition (such that $(2 y_i \cdot y_j -1)$ equals 1 if the labels are the same, otherwise $-1$).
The geometric difference between the classical and quantum kernel, $g_{CQ}$, is defined as...

\begin{equation}
    \centering
    g_{CQ} = \sqrt{\Vert\sqrt{K_Q}(K_C)^{-1}\sqrt{K_Q}\Vert_\infty}
\end{equation}
...where $K_C$ is the classical kernel matrix, $K_Q$ is the quantum kernel matrix, and $\Vert \hspace{5pt} \Vert_\infty$ is the spectral norm \cite{huang_power_2021}. The matrix square roots and inverses can be efficiently computed using the singular value decomposition. See the Supplemental Information (SI) for more details. 

The model complexity is used to define an upper bound on the model's expected error as a function of the number of training data points. For a model $h(x)$ trying to predict some true labels $f(x)$, the expected value of the absolute difference between the model predictions and the true labels over the input domain $\mathbb{D}$ is...
\begin{equation}
    E_{x\in \mathbb{D}}(\vert h(x)-f(x) \vert)\leq c\sqrt{\frac{s_K(N)}{N}}
\end{equation}
...for some constant $c>0$ \hspace{5pt}\cite{huang_power_2021}. This upper bound on the expected error decreases as more data is collected and increases with increasing model complexity, so the model complexity should be as small as possible to reduce the expected error. If the classical model complexity is much higher than the quantum model complexity, than the quantum model might have lower expected error when $N$ is small (i.e., when there is very little training data). 

The geometric difference is used to define another upper bound, this time on the classical model complexity. The geometric difference and model complexities obey the following relation:
\begin{equation}
    s_C \leq g_{CQ}^2 s_Q
\end{equation}
...where $s_C$ and $s_Q$ are the classical kernel and quantum kernel model complexities, respectively \cite{huang_power_2021}. In order for the classical model complexity to be much higher than the quantum model complexity, the geometric difference must be large. 

First, to predict whether a quantum kernel might have an advantage over classical kernels for any dataset, one uses the geometric difference. If the geometric difference is small, no advantage exists for that quantum kernel because the largest classical model complexity will be close to the quantum model complexity, regardless of dataset \cite{huang_power_2021}. However, if the geometric difference is large, there exists a dataset for which the classical model complexity (and thus the upper bound on the expected error) is much higher than the quantum model \cite{huang_power_2021}. 

Second, to predict whether a quantum kernel might have an advantage over classical kernels for a specific dataset, one uses the model complexities. If the model complexities are similar, the expected error upper bounds are also similar, and the classical model can learn the dataset as easily as the quantum model \cite{huang_power_2021}. However, if the classical model complexity is much higher than the quantum model complexity, then the upper bound on the expected error of the classical model is also much higher, and it's possible that the quantum model learns the dataset faster than the classical model \cite{huang_power_2021}. 

In addition to Huang et al.'s analysis \cite{huang_power_2021}, we empirically tested the quantum kernel on a supervised learning task adjacent to the typical autonomous workflow described above. In the above workflow, an XRD kernel function is used for the unsupervised task of clustering the XRD patterns into groups, while Huang et al.'s analysis applies to the supervised learning task of assigning labels to new data based on labeled training data. The supervised extrapolation step of the above workflow uses a composition kernel function, not an XRD kernel function. So, instead of using the quantum kernel for clustering XRD patterns, as in the above workflow, we studied a supervised learning task similar to tasks in the autonomous workflow: we trained a Gaussian process model to extrapolate the phase labels from human-expert-labeled XRD patterns to unlabeled XRD patterns. This task is a useful proxy for understanding the utility of the quantum kernels for autonomous materials science because we expect that a model which requires less training data to achieve a certain accuracy threshold for this task would also require less training data to achieve a certain uncertainty threshold in an active learning setting. 

To evaluate the performance of the kernels, we measured the subset accuracy of the Gaussian process classifier as a function of the number of training data points. For each training set size, $N$, we chose 20 subsets of size $N$ from the 20 experimental XRD  data points for training. The subsets were chosen uniformly at random without regard for the true labels (except for $N=19$ training data points, where we used all 20 subsets of size 19). We then calculated the number of correct test labels for each training set. In the parlance of FSL, we evaluated each kernel's 5 way $n$ shot accuracy, for $n=1$ to 4 depending on the training set size and random samples. We then compared the relative accuracy of the quantum and classical kernels, expressed as the ratio of their accuracies. This is analogous to relative risk, for which confidence intervals can be computed using the Katz method (Method C in ref. \hspace{0 pt} \cite{katz_obtaining_1978}) as implemented in SciPy, wherein the log of the ratio is taken to be normally distributed.

\section{Results and Discussion}

\subsection{Kernel Matrix Values}

\begin{figure*}[!b]
    \includegraphics[width=17cm]{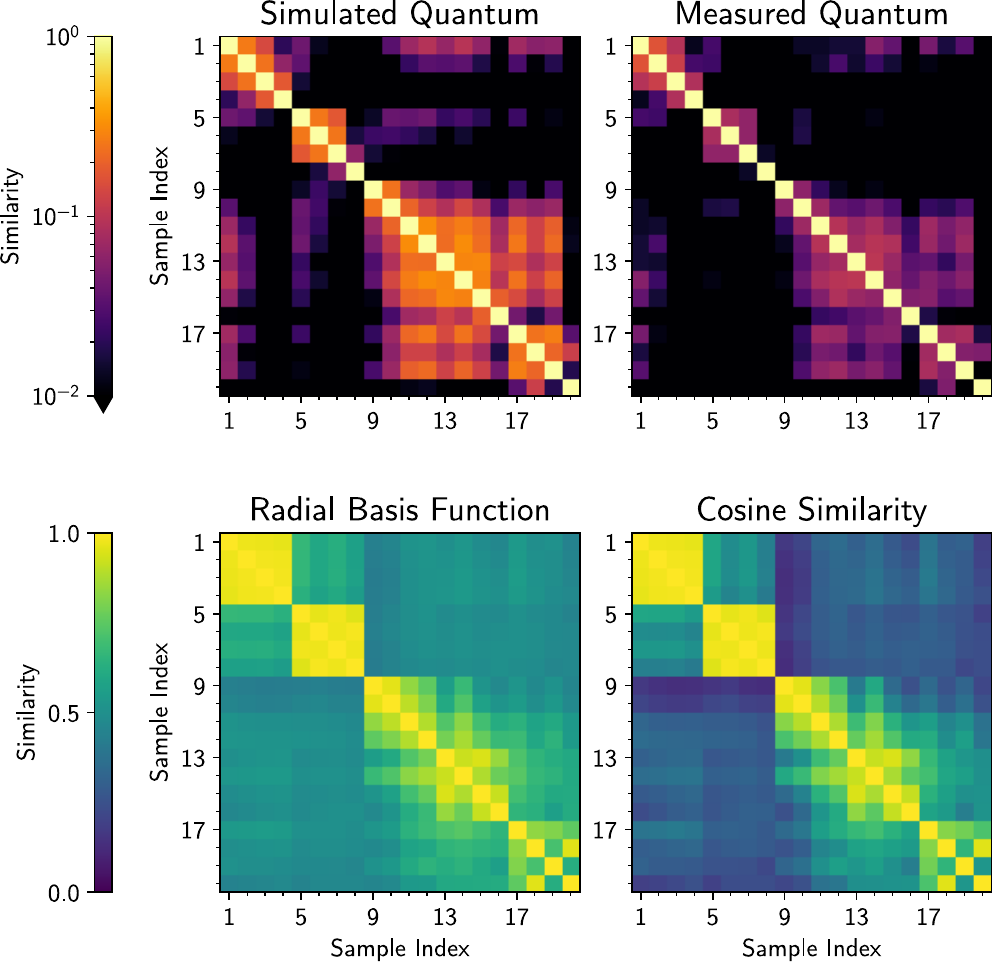}
    \caption{
        Comparison of the quantum and classical kernel matrices. The axis values are sample indices from Figure \ref{fig:active_learning}a. Quantum kernel matrices are shown with a logarithmic colormap to show the fine details not picked up by the classical kernels, while the classical kernels are shown with a linear colormap. See SI for all kernels in both logarithmic and linear scales.
    }
    \label{fig:kernel_matrices}
\end{figure*}

The quantum kernel matrix displays relationships between XRD patterns which weren't found by either of the classical kernels. Figure \ref{fig:kernel_matrices} shows the kernel matrices for all 20 XRD patterns. Figure \ref{fig:kernel_matrices} (a) and (b) show the quantum kernel matrices as simulated on a classical computer and measured on the IonQ Aria trapped ion quantum computer, respectively. Figure \ref{fig:kernel_matrices} (c) and (d) show the classical kernel matrices used for comparison. All the kernel matrices show that XRD patterns 1--4 and 5--8 form distinct groups which are more similar to each other than to other patterns and that patterns 9--12, 13--16, and 17--20 also form groups of 4 but are more interrelated. The quantum kernel found some faint inter-group similarities that weren't found by the classical kernels: patterns 1, 2, 5, and 6 are similar to several patterns outside their groups of 4. The quantum kernels are shown with a logarithmic scale to show these faint relationships, while the classical kernels are shown with a linear scale because they don't have any small values $< 0.1$. For comparison, all kernels in both linear and log scales are available in the SI. 

The relationships found by the quantum kernel and missed by the classical kernels are between XRD patterns with generally low intensity. Refer to Figure \ref{fig:active_learning} (b) to see the XRD patterns. Groups 1--4 and 5--8 both correspond to a body centered cubic Fe phase, but with significant peak shifting due to composition-dependent crystal lattice parameter changes \cite{long_rapid_2009}. However, in XRD patterns 1, 2, 5, and 6, the largest peak has a lower maximum intensity compared to other patterns in their groups. Similarly, XRD patterns 11--15 and 17--19 have no large peaks. 

These relationships between low intensity XRD patterns were missed by the classical kernels because they are magnitude-invariant, while the quantum kernel is magnitude-dependent. The classical kernels we used for comparison are defined in terms of normalized vectors, so they can't compare relative XRD intensities. The same is true for typical (but not all) XRD similarity measures, such as the Pearson's correlation coefficient. The output of the feature map circuit, however, does depend on the magnitude of the input data point, so the quantum kernel can compare the relative intensities of two XRD patterns. As a result, the quantum kernel identifies XRD patterns with generally low intensity as similar.

The kernel values as measured on the quantum computer are similar to the simulated values, except for a systematic decrease in kernel values due to additional hardware noise. As the calculation proceeds, errors accumulate and the two copies of the feature map circuit are less likely to cancel out, leading to decreased kernel values. Some of the fine details in the measured quantum kernel are sufficiently reduced that they are no longer detected because of the finite sampling used to calculate the kernel values.

\makeatletter
\close@column@grid
\onecolumngrid
\clearpage
\makeatother
\twocolumngrid

\clearpage

\subsection{Model Complexities and Performance} 

\begin{table}[thb]
    \begin{tabular}{c|cc}
        Kernel            & \begin{tabular}[c]{@{}c@{}}Geometric\\ Difference\end{tabular} & \begin{tabular}[c]{@{}c@{}}Model\\ Complexity\end{tabular} \\
        \hline
        Simulated Quantum     & 10.74 & 19.44 \\
        Measured Quantum      & 10.92 & 20.27 \\
        Cosine Similarity     &       & 36.85 \\
        Radial Basis Function &       & 37.20 \\
    \end{tabular}
    \caption{The results of the geometric difference and model complexity analyses. We calculated the geometric difference values with respect to both the classical kernels and show the worst case results (the Radial Basis Function kernel).}
    \label{tab:gd_results}
\end{table}

The geometric difference and model complexity calculations, shown in Table 1, suggest that the quantum kernel model could require less training data than the classical kernel models on the XRD dataset. The geometric difference between the classical and quantum kernels is greater than $\approx\sqrt{N}$, which indicates that there exists a dataset for which the classical model complexity is higher than the quantum model complexity. Indeed, the classical model complexity for our XRD dataset is higher than the quantum model complexity, which means that the upper bound on the expected error is higher for the classical kernel model. The analysis therefore suggests that it is possible that the quantum kernel model could have better performance with limited training data.

\begin{figure*}[!b]
    \includegraphics[width=17cm]{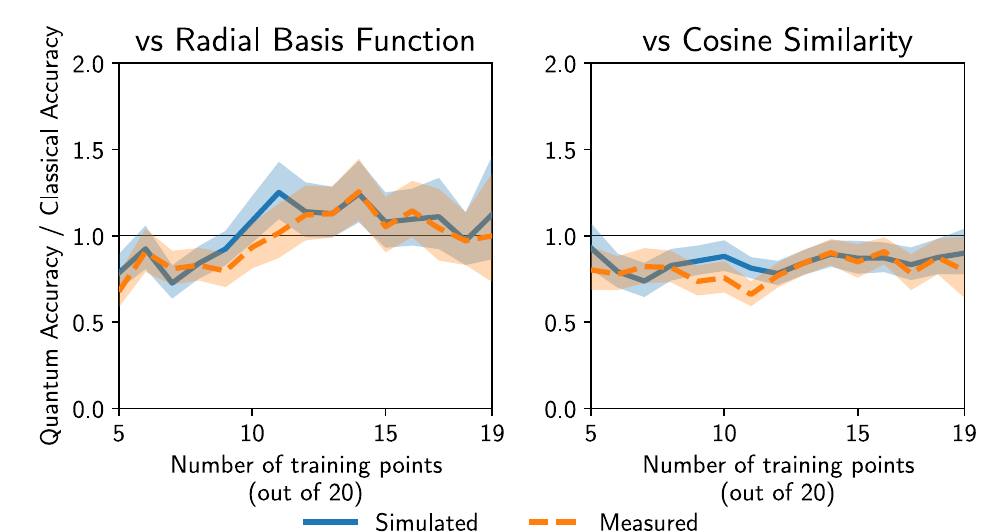}
    \caption{The relative performance of the quantum and classical kernels on the supervised learning task. Both plots show the accuracy of the Gaussian process classifier using the quantum kernel (simulated and measured) relative to one of the classical kernels. The left plot is relative to the radial basis function kernel and the right plot is relative to the cosine similarity kernel. The shaded regions indicate 95 \% confidence intervals.}
    \label{fig:kernel_performance}
\end{figure*}

Figure \ref{fig:kernel_performance} shows the performance of the Gaussian process model using the quantum kernel relative to the classical kernels as a function of the training set size. See the SI for separate kernel accuracies. The shaded regions indicate 95 \% confidence intervals. The confidence intervals expand as the number of training data points increase because there are fewer testing data points (out of the total 20 data points) with which to evaluate the model performance, decreasing the sample size. Comparing the performance of the simulated and measured quantum kernel models, the performance of the measured quantum kernel model is slightly lower for some training set sizes, indicating that reducing hardware noise could improve the performance of the quantum kernel model.

The empirical results show that the quantum kernel model requires less training data than only one of the two classical kernel models within a certain range of training set sizes. Between about 10 and 15 training data points, the quantum kernel model outperforms the radial basis function classical kernel model, until about 15 and 19 training data points, where the classical kernel has enough data to achieve similar performance. With less than 10 training data points, neither the quantum nor radial basis function kernel has much data, and the radial basis function kernel happens to perform better. The cosine similarity, despite having a kernel matrix not unlike the radial basis function kernel, outperforms all other kernels for this dataset.

\makeatletter
\close@column@grid
\onecolumngrid
\clearpage
\makeatother
\twocolumngrid

The model complexity and performance depend on the combination of dataset and kernel function. For example, the quantum model complexity is less than the classical model complexities, even though the quantum kernel matrix captured more nuanced relationships in the data because of its magnitude dependence. Also, the cosine similarity outperformed all other methods for this dataset despite its magnitude invariance.

\begin{figure*}[!b]
    \centering
    \includegraphics[width=17cm]{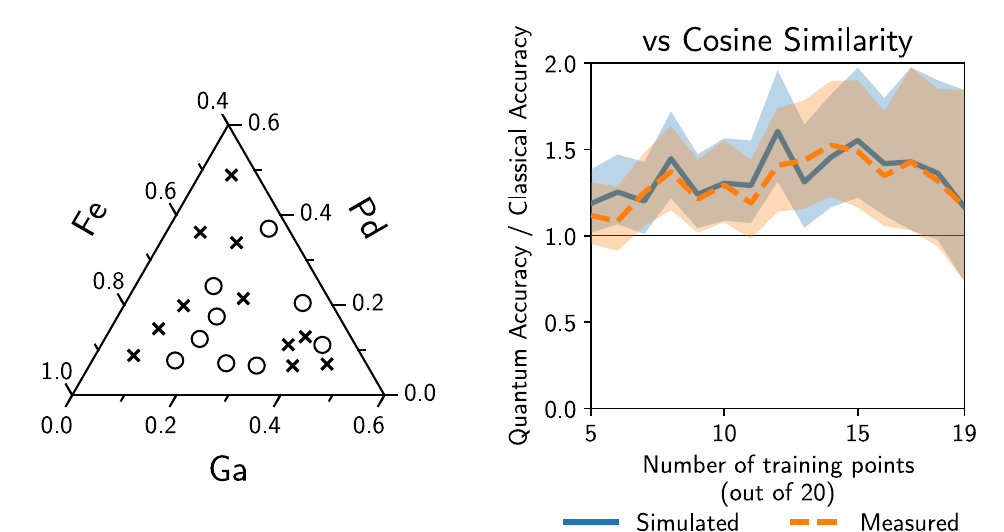}
    \caption{The relative performance of the quantum and classical kernels on the engineered labels. The ternary plot on the left shows the engineered labels of the XRD patterns as a function of their composition. Like Figure \ref{fig:kernel_performance}, the plot on the right shows the accuracy of the Gaussian process classifier using the quantum kernel relative to the performance the cosine similarity. Again, shaded regions indicate 95 \% confidence intervals.} 
    \label{fig:engineered_kernel_performance}
\end{figure*}

To further illustrate how sensitive the performance of the models is to their suitability for a particular problem, we compared the performance of the quantum kernel and cosine similarity on arbitrary, engineered labels constructed to minimize the performance of the cosine similarity. The results of this comparison are shown in Figure \ref{fig:engineered_kernel_performance}. Drawing on the work of Huang et al.\cite{huang_power_2021} again, these binary labels are constructed to maximize the model complexity of the cosine similarity kernel model. For details on the constructions of these labels, see the SI. With these labels, the cosine similarity, which previously outperformed all other models, does much worse than the quantum kernel. See the SI for the accuracy on the engineered data for each kernel separately. 

These results, and the work of Huang et al. \cite{huang_power_2021}, can be understood in terms of model bias and its relationship to model complexity. In this context, the bias of a model is the difference between the expected value of the model's prediction and the true value it's trying to predict \cite{mitchell_machine_1997}. As the complexity of a model increases, it can make more nuanced predictions, and thus its bias decreases \cite{luxburg_statistical_2008}. However, as Huang et al. showed, increasing the model complexity also increases the amount of data required to train the model \cite{huang_power_2021}. The data-limited regimes of materials science thus reward simple, low-complexity models. Simple models, however, are necessarily inflexible and must be carefully chosen to match the problem at hand. 

With this understanding, our results suggest that quantum kernel machine learning models can outperform classical kernel models, but the quantum kernel needs to have an appropriate inductive bias. The inductive bias of a machine learning model is the set of assumptions that the model uses to make predictions \cite{mitchell_machine_1997}. The quantum kernel outperforms the radial basis function kernel once enough data is collected because it has lower model complexity and thus requires less training data. However, the cosine kernel outperforms both other kernels because its inductive bias is a better fit for this classification task. This indicates that a quantum kernel with low model complexity and appropriate inductive bias constructed specifically to discriminate XRD patterns would have a quantum advantage and outperform classical kernels.

Such a quantum kernel with appropriate inductive bias might be found using the FSL technique of metric learning. The goal of metric learning is to learn a distance or similarity function which can then be used by a simple machine learning model such as k-nearest neighbors \cite{parnami_learning_2022, kulis_metric_2013}. Metric learning techniques could be used to identify or train feature map quantum circuits which produce quantum kernels performant on XRD classification tasks.

\makeatletter
\close@column@grid
\onecolumngrid
\clearpage
\makeatother
\twocolumngrid

Our results should be viewed in the broader context of problem-aware quantum models, where the circuit ansatz is designed to respect known problem structure, as in permutation-equivariant circuits and quantum convolutional architectures \cite{skolik_equivariant_2023, chinzei_splitting_2024}. In contrast, our XRD task does not exhibit an obvious global permutation or translational symmetry, so here we restrict attention to a single, generic feature map and analyze its behavior through the model-complexity framework. Systematically exploring symmetry-inspired and task-specific choices of $U$ for diffraction data is an important direction for future work.

\section{Conclusion}

In this work, we demonstrated that autonomous materials discovery provides a compelling potential use case for quantum kernel machine learning. Quantum machine learning holds exciting potential for advancing artificial intelligence frameworks, with its success depending critically on aligning the right quantum models with suitable task–data combinations \cite{huang_power_2021}. Building on the theoretical framework of Huang et al. \cite{huang_power_2021}, we applied quantum kernels to a real-world supervised x-ray diffraction classification task. We characterized both quantum and classical kernel models using their respective model complexities and performed experiments which demonstrated that quantum kernel models can achieve similar performance to classical models while requiring less training data. 

Looking ahead, there are significant opportunities for further exploration in quantum kernel design and integration into autonomous materials science. Our results show that by carefully choosing a combination of problem and quantum kernel, it's possible to achieve quantum advantage for classification of x-ray diffraction patterns. More work is needed to better understand this relationship and develop quantum kernels which are well suited to problems of interest in diffraction pattern analysis. As quantum hardware advances, more complex and expressive feature map circuits will become viable, enabling even greater flexibility in tailoring quantum kernel models to the unique characteristics of each dataset. More work is also needed to explore how effectively these quantum kernels can guide active learning in the autonomous setting. This progress opens the door to applying quantum kernel models to increasingly sophisticated datasets, reinforcing their promise as a powerful tool in artificial intelligence-driven materials discovery and beyond.

\newpage

\section*{Supplemental Information}

The Supplemental Information contains the code used to calculate the model complexities and geometric differences, as well as some extra re-plots of the data in this work.

\begin{acknowledgments}

This work was supported by an IonQ seed grant and partly supported by NIST grant No. 60NANB19D027.

\end{acknowledgments}

\section*{NIST Disclaimer}

Certain commercial equipment, instruments, or materials are identified in this report in order to specify the experimental procedure adequately. Such identification is not intended to imply recommendation or endorsement by the National Institute of Standards and Technology, nor is it intended to imply that the materials or equipment identified are necessarily the best available for the purpose.

\section*{Author Declarations}

\subsection*{Conflict of Interest Statement}

The authors have no conflicts to disclose.

\subsection*{Author Contributions}

In order of size of contribution, beginning with largest contribution; 
\textbf{Conceptualization}: equal; 
\textbf{Formal Analysis}: F.A.; 
\textbf{Funding Acquisition}: I.T.; 
\textbf{Investigation}: F.A.; 
\textbf{Methodology}: equal; 
\textbf{Software}: F.A., A.G.K.; 
\textbf{Supervision}: I.T.; 
\textbf{Visualization}: F.A.; 
\textbf{Writing / Original Draft Preparation}: F.A.; 
\textbf{Writing / Review \& Editing}: equal

\section*{Data Availability Statement}

The data and code that support the findings of this study are openly available on GitHub: \url{https://github.com/fadams-umd/qkml_for_autonomous_matsci}

\clearpage

\section*{References}

\bibliography{qkml_references.bib}

\end{document}


\title{Quantum Kernel Machine Learning for Autonomous Materials Science:\newline Supplemental Information}

\maketitle 

\section{Calculation of Model Complexity and Geometric Difference}

In order to calculate the geometric difference and model complexity using the equations in ref.\hspace{5pt}\cite{huang_power_2021}, we used the Singular Value Decomposition as implemented in NumPy:

\begin{minted}{python}
    import numpy as np

    # calculate the SVD of the classical and quantum matrices
    cU, cS, cVh = np.linalg.svd(classical_kernel_matrix, hermitian=True)
    qU, qS, qVh = np.linalg.svd(quantum_kernel_matrix, hermitian=True)

    # use the SVD to calculate the square root and inverse matrices
    sqrt_qK = qU @ np.diag(np.sqrt(qS)) @ qVh
    
    inv_cK = cVh.T @ np.diag(cS**(-1)) @ cU.T

    # calculate the singular values of the prescribed matrix product
    tS = np.linalg.svd(sqrt_qK @ inv_cK @ sqrt_qK, compute_uv=False)

    # calculate the geometric difference from the singular values
    g_cq = np.sqrt(np.max(tS))

    

    # calculate the other inverse matrix
    inv_qK = qVh.T @ np.diag(qS**(-1)) @ qU.T
    
    # calculate the model complexities
    s_C = 0
    s_Q = 0

    for i in range(inv_cK.shape[0]):
        for j in range(inv_cK.shape[1]):
            s_C += inv_cK[i, j] * (1 if true_labels[i] == true_labels[j] else -1)
            s_Q += inv_qK[i, j] * (1 if true_labels[i] == true_labels[j] else -1)
\end{minted}

\section{Construction of Engineered Labels}

Similarly, we use the construction from the supplemental information of ref.\hspace{5pt}\cite{huang_power_2021} to create the engineered labels.

\begin{minted}{python}
    # calculate the eigenvalues and eigenvectors of the prescribed matrix product
    eig_vals, eig_vecs = np.linalg.eig(sqrt_qK @ inv_cK @ sqrt_qK)

    # threshold the values in the eigenvector with the largest eigenvalue to create the labels
    engineered_labels = np.array([0 if x < 0 else 1 for x in sqrt_qK @ eig_vecs[:, 0]])
\end{minted}

\section{References}
\bibliography{qkml_references}

\section{Kernel Matrices With All Logarithmic Scales}

\begin{figure*}[h]
    \includegraphics[width=17cm]{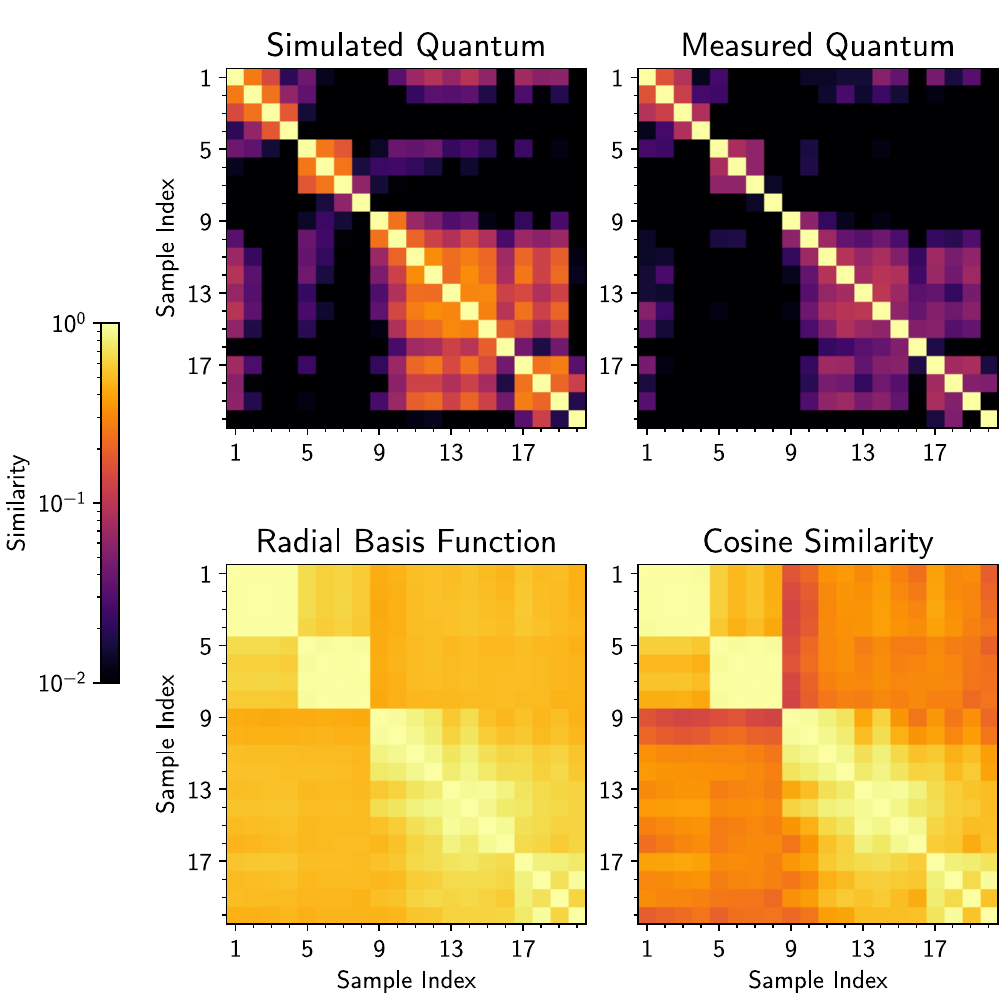}
    \caption{A re-plot of the kernel matrices, all with the logarithmic scale.}
\end{figure*}

\clearpage

\section{Kernel Matrices With All Linear Scales}

\begin{figure*}[h]
    \includegraphics[width=17cm]{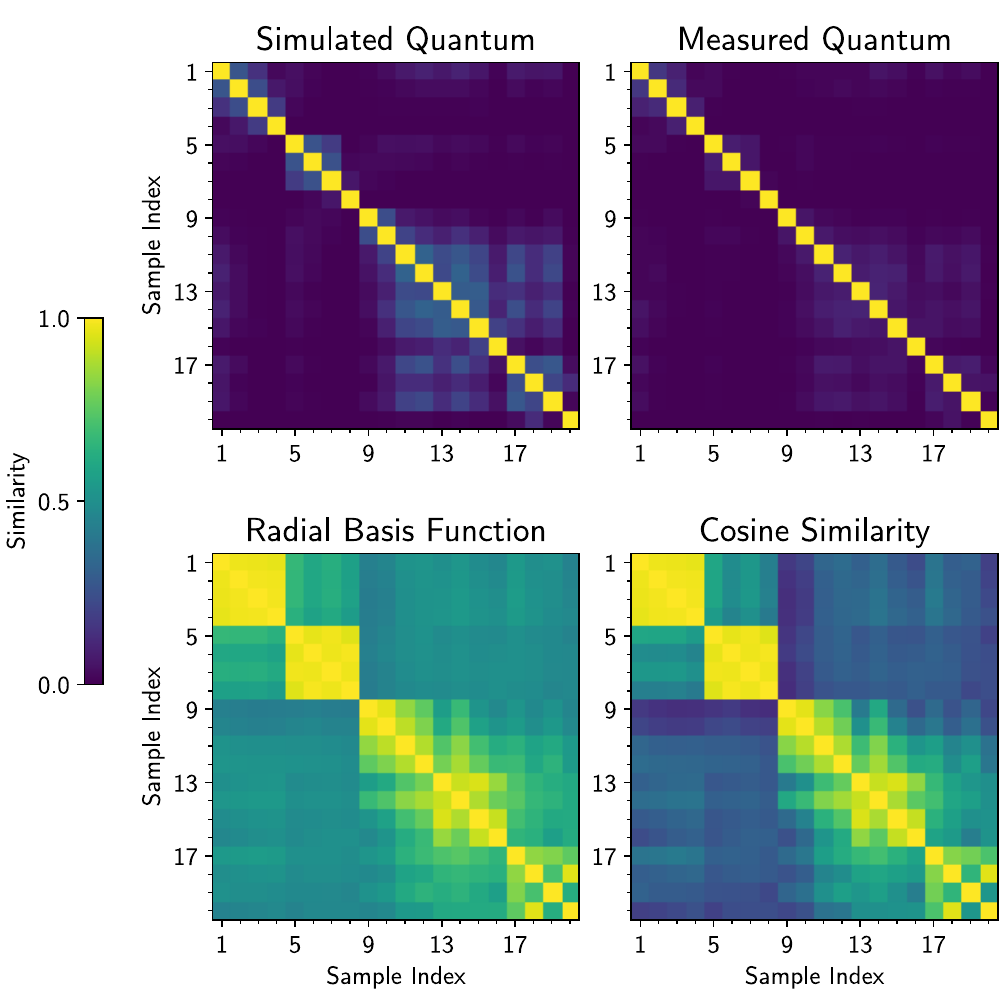}
    \caption{A re-plot of the kernel matrices, all with the linear scale.}
\end{figure*}

\clearpage

\section{Kernel Accuracies on True Labels}

\begin{figure*}[h]
    \includegraphics[width=17cm]{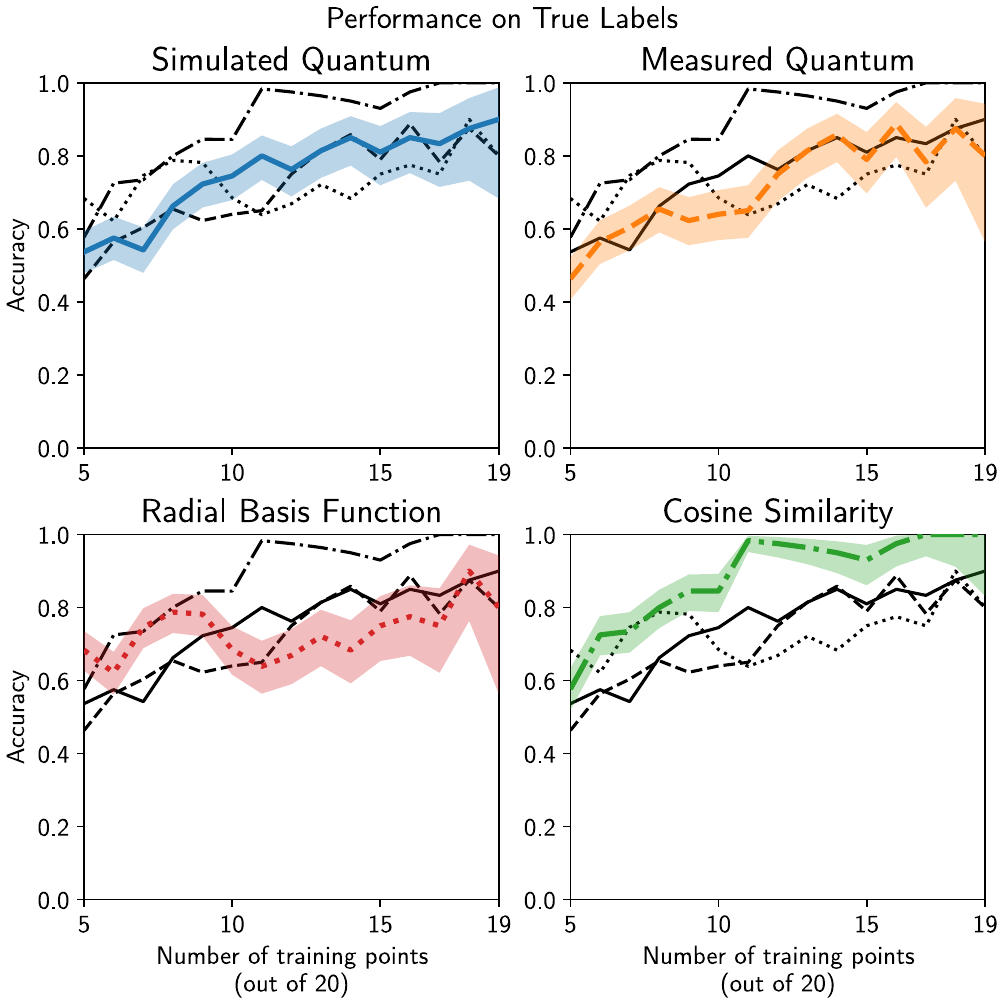}
    \caption{The absolute accuracy of each of the kernels, as opposed to the relative accuracy of the quantum kernel, on the material science supervised learning task.}
\end{figure*}

\clearpage

\section{Kernel Accuracies on Engineering Labels}

\begin{figure*}[h]
    \includegraphics[width=17cm]{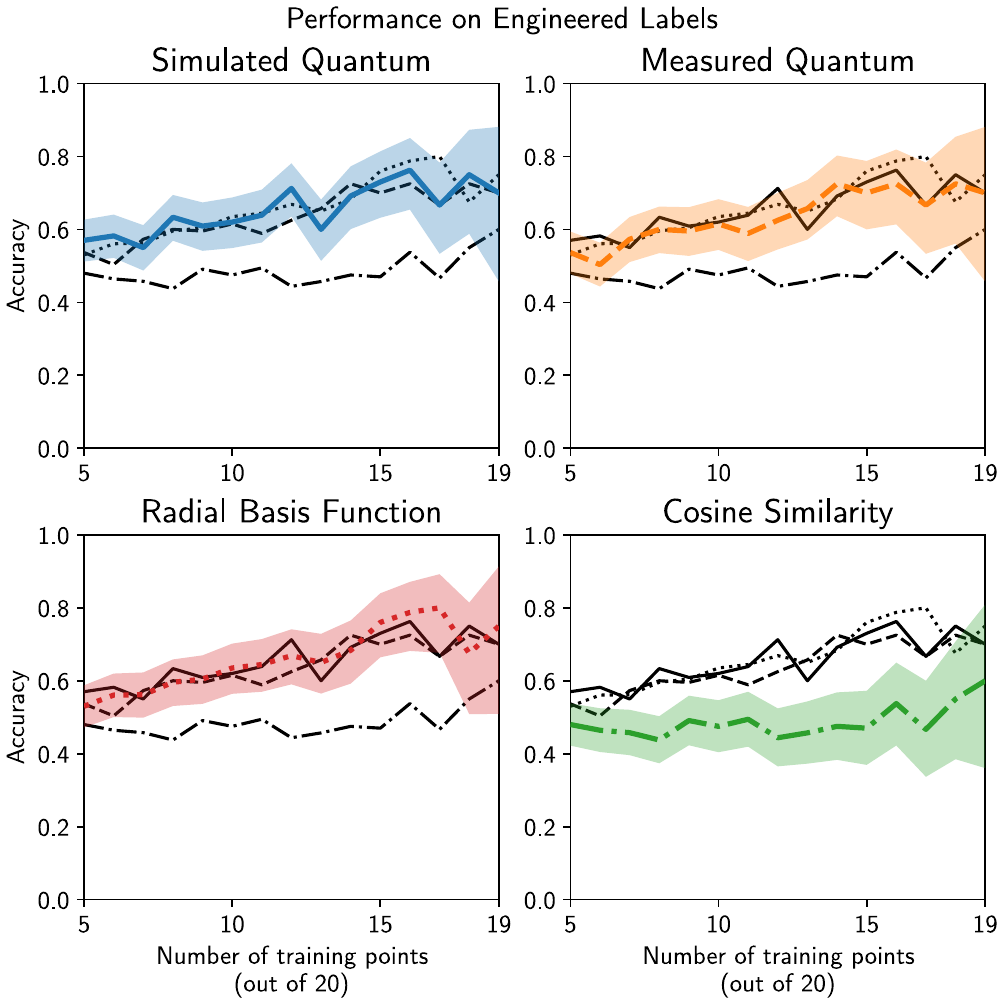}
    \caption{The absolute accuracy of each of the kernels, as opposed to the relative accuracy of the quantum kernel, on the engineered label supervised learning task.}
\end{figure*}

\clearpage

\section{Confusion Matrices for Select Training Set Sizes}

\begin{figure*}[h]
    \includegraphics[width=17cm]{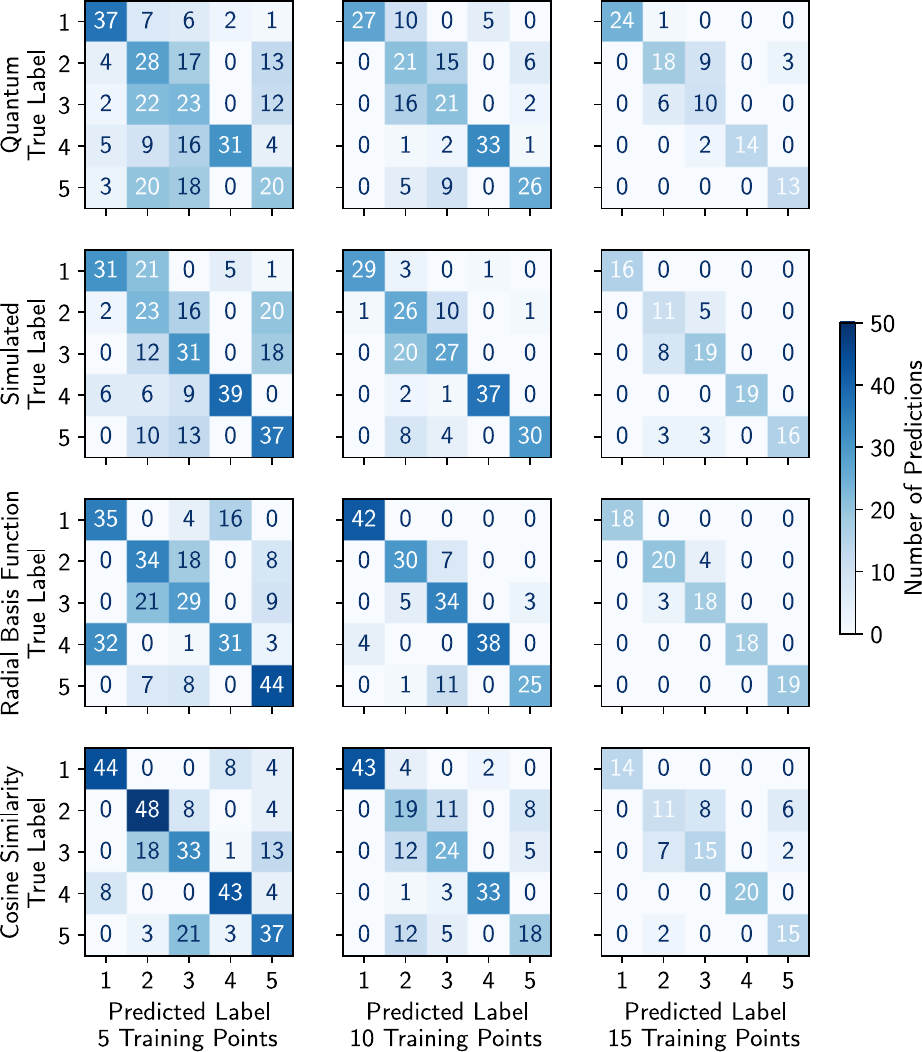}
    \caption{Some confusion matrices for the materials science supervised learning task. The above confusion matrices are the sum of the 20 confusion matrices for each random training set sample. }
\end{figure*}